\documentclass[
    ,final            
  ]
  {aipproc}

\layoutstyle{6x9}

\begin{document}

\title{Deeply--Virtual Compton Scattering \\ on Deuterium and Neon at HERMES}

\author{F. Ellinghaus$^*$\footnote{E-mail: Frank.Ellinghaus@desy.de} , 
R. Shanidze$^\dagger$ and J. Volmer$^*$ \\
 \small {(On behalf of the HERMES Collaboration)}}
{address={$^*$DESY Zeuthen, Platanenallee 6, 15738 Zeuthen, Germany \\
$^\dagger$Universit\"at Erlangen-N\"urnberg, Erwin--Rommel--Str. 1, 
91058 Erlangen, Germany}}

\begin{abstract}
We report the first observation of azimuthal beam--spin 
asymmetries in hard electroproduction of real photons off nuclei.
Attributed to the interference between the Bethe--Heitler 
process and the deeply--virtual Compton scattering  
process, the asymmetry gives access to the latter at the amplitude level.
This process appears to be the theoretically cleanest way
to access generalized parton distributions.
The data presented here have been accumulated by the 
HERMES experiment at DESY,
scattering the HERA 27.6 GeV positron beam
off deuterium and neon gas targets. 
\end{abstract}

\maketitle


\section{Introduction}
Hard scattering processes, such as inclusive 
deeply--inelastic scattering (DIS), semi--inclusive DIS 
and 
hard exclusive scattering have an important property in common,
namely the possibility to separate
the exactly calculable perturbative parts of the reaction
from the non--perturbative parts. 
This factorization property is well established 
in the case of inclusive and semi--inclusive DIS and has been extensively
used to investigate the structure of the nucleon.
Only a few years ago, factorization theorems have been established
for some hard exclusive reactions, where the produced particle 
is e.g. a photon \cite {Rad97,Ji98,coll_freund}. 
Their description in the theoretical framework of generalized
parton distributions (GPDs) \cite{Mue94,Ji97,Rad97}
takes into account the dynamical correlations between partons
of different momenta in the nucleon.
The ordinary parton distribution functions and form factors 
turn out to be the limiting cases and moments of GPDs, respectively.
Of particular interest is the second moment of two 
unpolarized quark GPDs, which for the first time offers
a possibility to determine the total angular momentum carried
by the quarks in the nucleon \cite{Ji97}.
Recent theoretical ideas indicate that 
GDPs are able to 
describe correlations between the longitudinal
and transverse structure of the nucleon \cite{Bur,Die02}.
For the case of coherent hard exclusive processes on nuclei it was 
pointed out very recently \cite{maxim} that
information about the distribution of energy, pressure, and shear forces
inside nucleons and nuclei become accessible. 

\section{Deeply--Virtual Compton Scattering}
In deeply--virtual Compton scattering (DVCS)
a photon with large virtuality $Q^2$ is absorbed by a parton
inside the nucleon and a real photon is produced.
This process is considered to be the 
theoretically cleanest way to access GPDs.
The DVCS cross section can be obtained through a measurement of
the exclusive photon production cross section after subtracting
the background from the Bethe--Heitler (BH) process 
which has an identical final state and is calculable exactly in QED.
First results on the DVCS cross section at high energies have
been published recently by H1 \cite {adloff} and ZEUS \cite {saull}. 
At the lower
energies of HERMES at DESY and CLAS at Jlab, the DVCS cross section is 
expected to be much smaller than the BH cross section and thus
a measurement with sufficient precision is not yet
feasible. However, the DVCS amplitudes 
are directly accessible through the interference between the DVCS
and BH processes. 
The leading--order and leading--twist interference term \cite{Die97}
\begin{equation} \label {Diehl}
I = \pm \, \frac {4 \, \sqrt 2}{ t \,  Q \, x_B} 
\frac{m_p \, e^6}{\sqrt {1-x_B}} \times 
\left [ \cos \phi \, \frac{1}{\sqrt {\epsilon (\epsilon - 1)}}
\, \mathrm{Re} \, \tilde M^{1,1} 
- P_l \, \sin \phi \, \sqrt { \frac {1 + \epsilon}{\epsilon}} 
\, \mathrm{Im} \, \tilde M^{1,1} \right ]
\end{equation}
depends on the charge and the polarization of the incident lepton, 
where +(-) in front of the expression corresponds to a negatively 
(positively) charged lepton with polarization $P_l$. 
Here $m_p$ represents the proton mass, $t$ the square of the four--momentum 
transfer to the target, $-Q^2$ the virtual--photon four--momentum squared,
$x_B$ the momentum fraction of the nucleon carried by the struck quark,
and $\epsilon$ is the polarization parameter of the virtual photon. 
The azimuthal angle $\phi$ is defined as the angle between the lepton 
scattering plane, i.e. the plane defined by the incoming and the 
outgoing lepton trajectories,
and the photon production plane made up by the virtual and real photons.
The linear combination of DVCS amplitudes $\tilde M^{1,1}$ 
can be expressed as a linear combination of GPDs convoluted
with hard scattering amplitudes.

Appropriate cross section asymmetries allow the separate access
to the real and imaginary part of $\tilde M^{1,1}$.
The beam--spin asymmetry (BSA)  
\begin{equation} \label {bca}
A_{LU} (\phi) = \frac {1}{< \left | P_l \right | >} \, 
\frac {\overrightarrow N (\phi) - \overleftarrow N (\phi)}
{\overrightarrow N (\phi) + \overleftarrow N (\phi)} \sim 
\sin \phi \, \mathrm{Im} \, \tilde M^{1,1}
\end{equation}
is proportional to the imaginary part of $\tilde M^{1,1}$,
where the average polarization of the beam 
is given by $< \left | P_l \right | >$ and $\overrightarrow N$
($\overleftarrow N$) represents the normalized
yield for positive (negative) beam helicity.
The subscripts $L$ and $U$ denote a longitudinally polarized
beam and an unpolarized target. 
Measurements of the BSA on the proton have already been carried 
out by HERMES \cite {HER_DVCS} and CLAS \cite{clas}.
The new preliminary result on the proton
based on HERMES data collected in 2000
is shown in the left panel of figure \ref {yield}.
The asymmetry indeed shows the expected $\sin \phi$ modulation.
Note that although 
the average kinematic values are slightly different compared 
to those from the already published BSA from the 1996/97 
running period \cite {HER_DVCS}, 
the results are consistent with each other.

Recently, the first measurement of the beam--charge asymmetry, 
accessing the real part of the same combination of DVCS amplitudes, 
has been carried out at HERMES \cite {ich} 
via the scattering of positron 
and electron beams off an unpolarized hydrogen target.

\section{DVCS on Nuclei}
The data presented in the following have been accumulated 
by the HERMES experiment \cite {HER_Spec} at DESY during the 2000 
running period. 
The HERA 27.6 GeV positron beam was scattered off 
polarized and unpolarized deuterium and unpolarized neon gas targets. 
Both the unpolarized and the spin--averaged polarized target data 
have been used in this analysis.
Selected events contained exactly one photon and one charged track 
identified as the scattered positron. 
The kinematical requirements were $Q^2 > 1$~GeV$^2$, 
$W^2 > 4$~GeV$^2$ and $\nu < 23$~GeV. Here $W$ denotes the 
photon--nucleon invariant mass and $\nu$ is the virtual--photon energy.
The angle between the real and the virtual photon was required to be 
within 2 and 70~mrad.
\begin{figure} 
\includegraphics[width=0.52\textwidth]{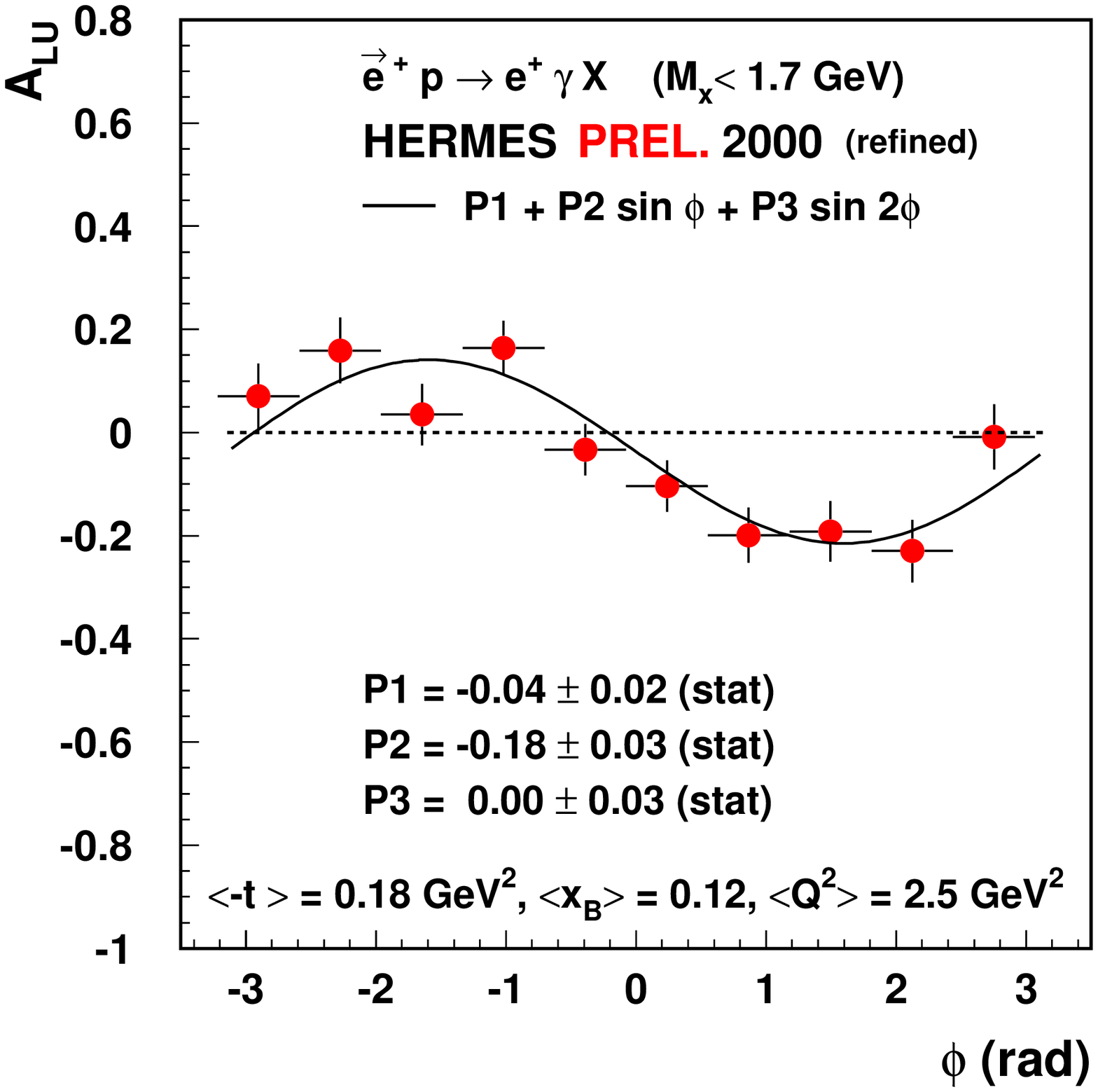} 
\includegraphics[width=0.52\textwidth]{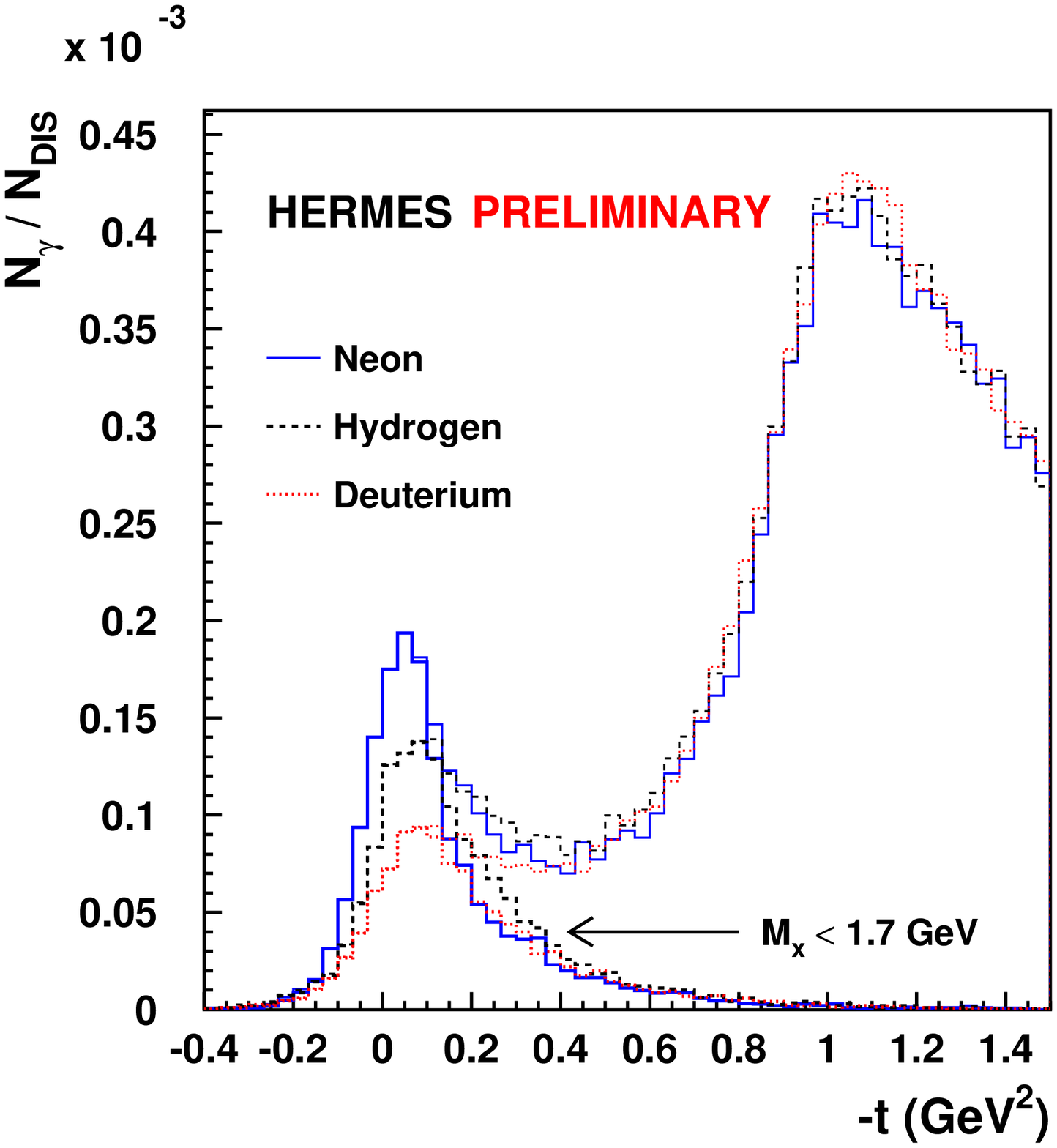}
\caption{Left panel: Beam--spin asymmetry $A_{LU}(\phi)$ for 
the hard exclusive electroproduction of photons off the proton. 
Exclusive events are defined through the missing mass constraint
$M_x < 1.7$~GeV. Right panel: $-t$ distribution of 
single-photon yields for neon, deuterium and hydrogen normalized 
to the number of DIS events. The exclusive events ($M_x < 1.7$~GeV) 
are shown separately.}
\label{yield}
\end{figure}
In the right panel of figure \ref{yield} the normalized 
yield of single photon events
is shown versus $-t$ for neon and deuterium in comparison to 
hydrogen. Note that negative values of $-t$ appear due to the finite
resolution of the spectrometer.
Since the recoiling nucleus is not detected in the HERMES
spectrometer, the missing mass  
$M_x = \sqrt {(q + P - k)^2}$ is calculated
from $q$, $P$ and $k$, the four--momenta of 
virtual photon, target nucleus and real photon,
respectively. For this analysis the target mass 
is set to the proton mass in order to keep the same $M_x < 1.7$~GeV
definition for exclusive events regardless of the target.
The cross section for the exclusive events is dominated by 
the BH contribution, i.e. when going from nucleon to nuclei
it increases with the square of the charge diminished by the 
form factor squared. This explains the differences
in the single--photon yield for the different targets at small 
values of $-t$, i.e. for exclusive events as shown in the right 
panel of figure~\ref {yield}.

\begin{figure} 
\includegraphics[width=0.52\textwidth]{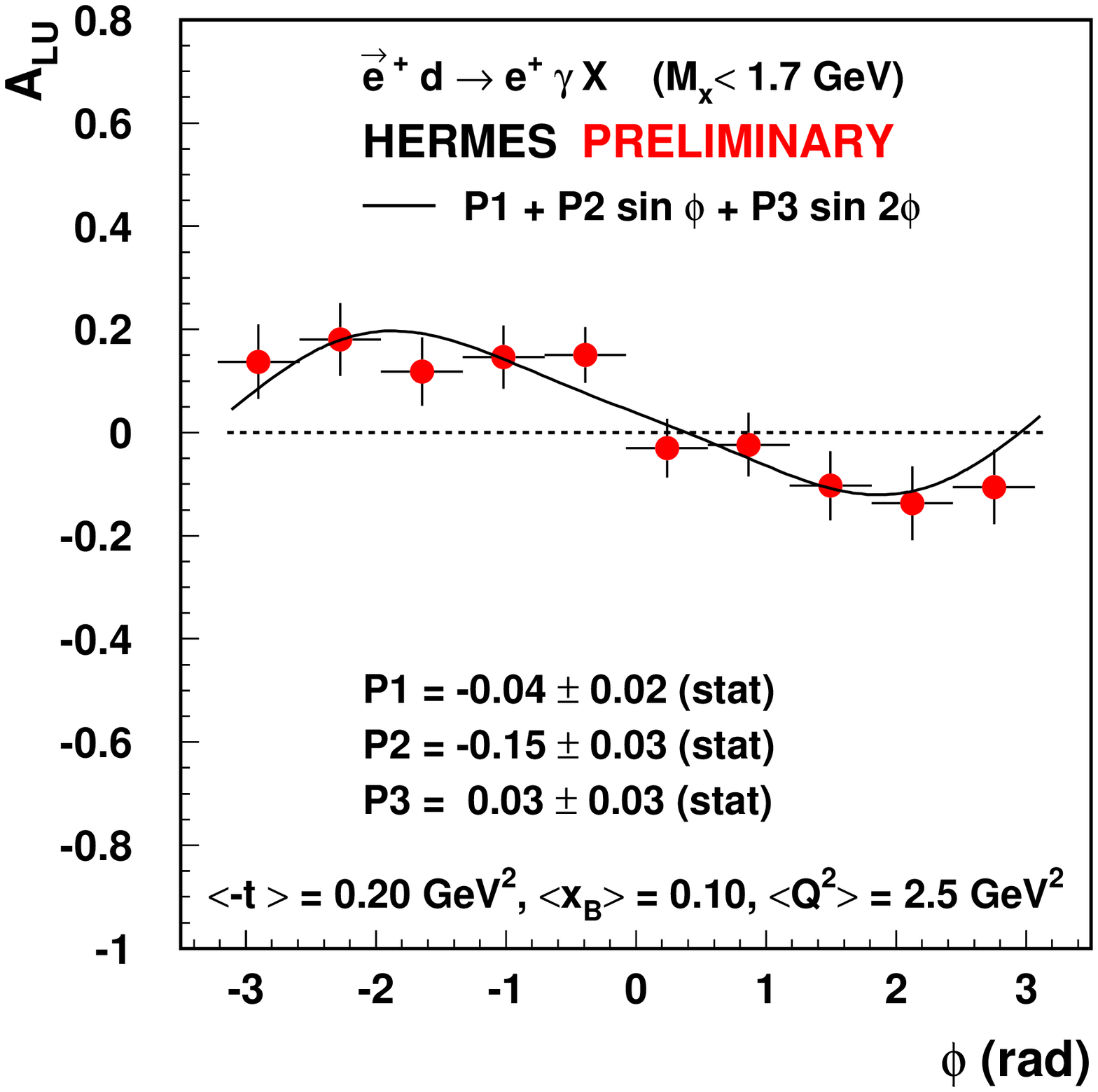}
\includegraphics[width=0.52\textwidth]{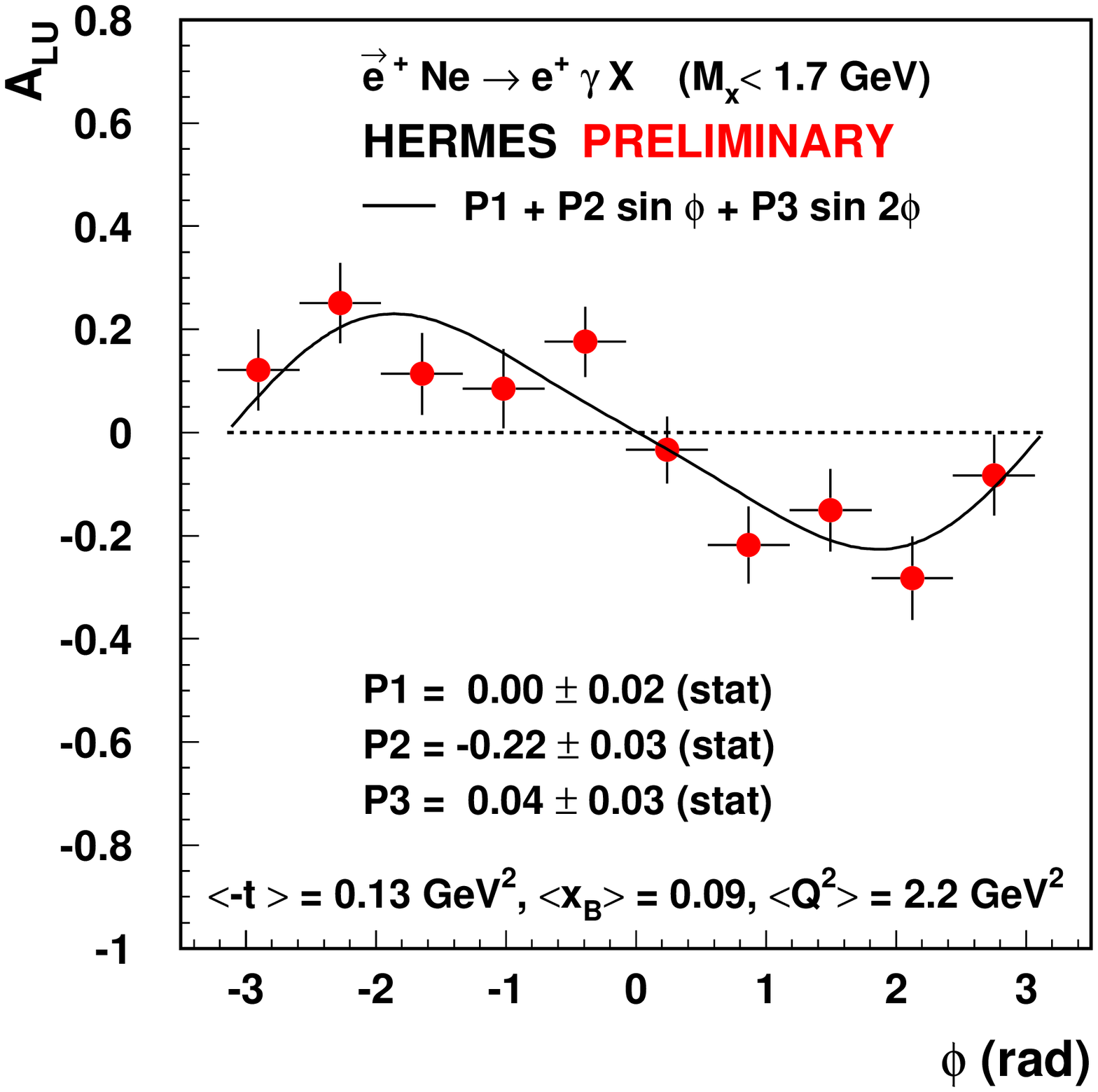}
\caption{Beam--spin asymmetries $A_{LU}(\phi)$ for 
the hard electroproduction of photons off deuterium (left panel)
and neon (right panel)
for events with a missing mass $M_x < 1.7$~GeV.}
\label{bsas}
\end{figure}

\begin{figure} 
\includegraphics[width=0.52\textwidth]{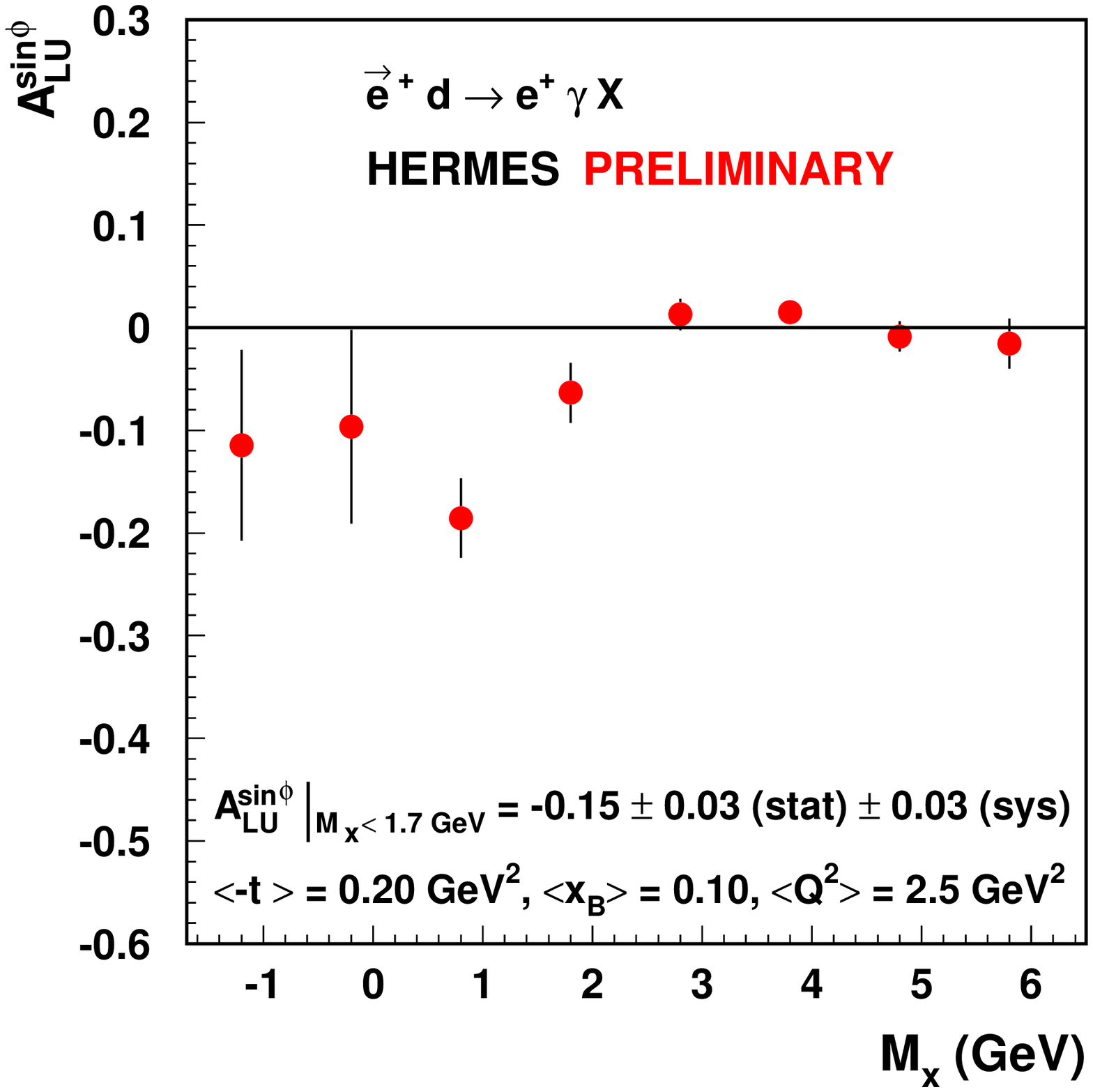}
\includegraphics[width=0.52\textwidth]{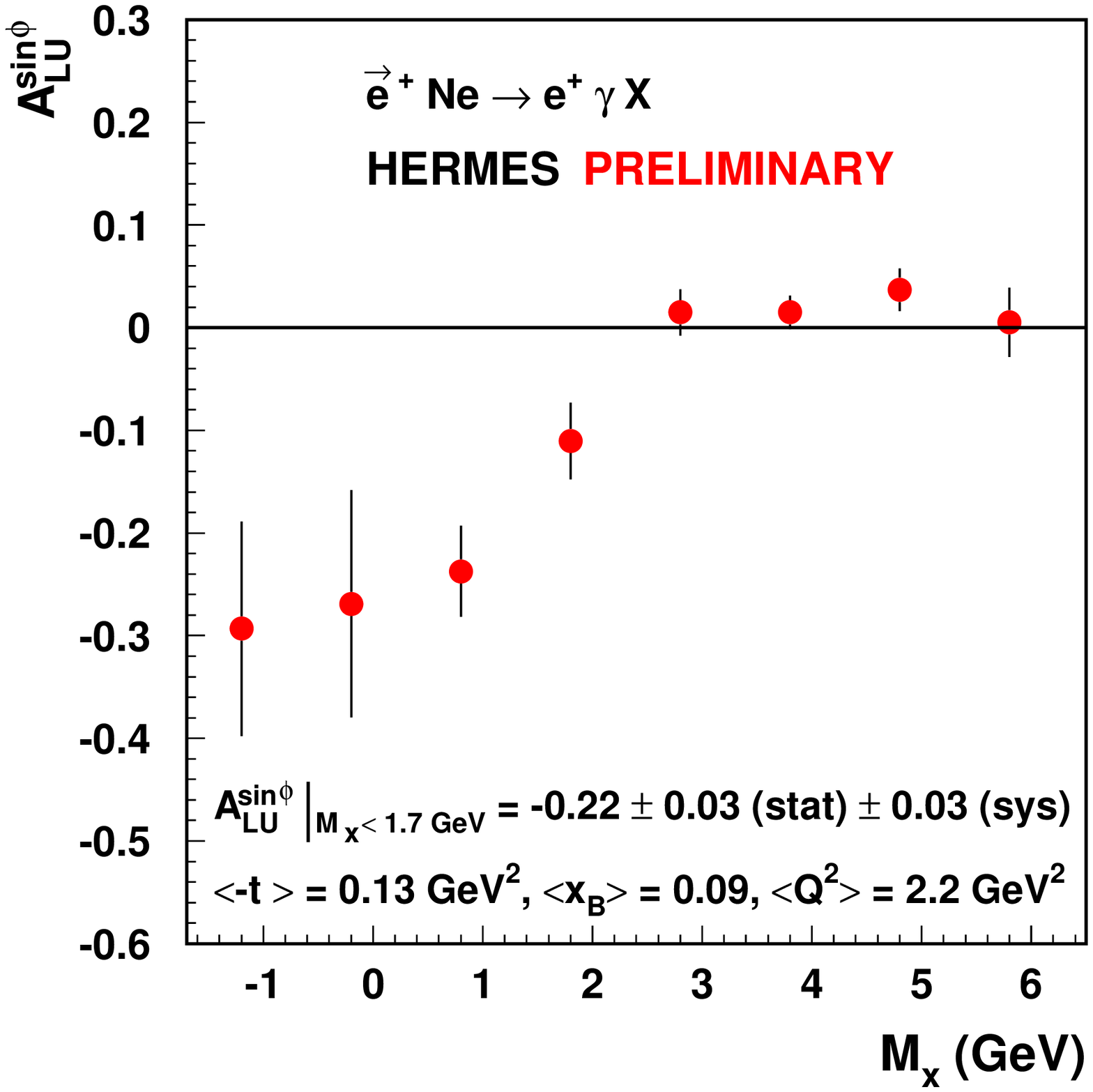}
\caption{$\sin \phi$--weighted moments $A_{LU}^{\sin\phi}$ for 
the hard electroproduction of photons off deuterium (left panel)
and neon (right panel) versus the missing mass $M_x$.}
\label{sins}
\end{figure}

In figure \ref{bsas} the azimuthal dependences of the 
BSAs on deuterium and neon are shown for events with a missing 
mass $M_x$ below 1.7~GeV.
At the given average kinematics as indicated in the plots,
the data exhibit the
expected $\sin \phi$ behavior represented by the fit to the function
$P_1 + P_2 \sin (\phi) + P_3 \sin (2\phi)$.
Note that $x_B$ is calculated using the proton mass, 
i.e. $x_B = Q^2 / 2 \, m_p \, \nu$. 
As an independent method the
$\sin \phi$--weighted moments    
\begin{equation}
A_{LU}^{\sin\phi} = \frac{2}{\overrightarrow N + \overleftarrow N}
\sum_{i=1}^{\overrightarrow N + \overleftarrow N} 
\frac {\sin\phi_i} {(P_l)_i} 
\end{equation}
are shown in figure \ref {sins} versus
the missing mass $M_x$.
The moments are non--zero only in the exclusive region and 
integrating $A_{LU}^{\sin\phi}$ up to $M_x < 1.7$~GeV yield 
the same results as the fits for the parameter $P_2$.
Note, that negative values of the missing mass are again 
a consequence of the
finite momentum resolution of the spectrometer,
in that case $M_x = - \sqrt {-M_x^2}$ was defined.
In contrast to the case of DVCS on the proton, DVCS on the deuteron 
has received little consideration in the literature 
\cite {berger,kir_muell,cano_pire}. 
The BSA on the deuteron is expected
to be slightly smaller \cite {kir_muell} than the one on the
proton.
This is in agreement with our results
when comparing the left panels 
in figure \ref {yield} and figure \ref {bsas},
where the BSAs on the proton and the deuteron, 
achieved in a similar average kinematic region, 
amount to $-0.18 \pm 0.03$~(stat) $\pm \, 0.03$~(sys) 
and $-0.15 \pm 0.03$~(stat) $\pm \, 0.03$~(sys), respectively.
However, present theoretical predictions assume 
$Q^2 \ge 4$~GeV$^2$ in order to avoid possible large 
target--mass corrections which have not yet
been calculated for spin--1 targets. In addition, since at
HERMES the recoiling nucleus is presently not
detected, the ratio of coherent to incoherent production cannot 
be directly inferred from the measurement.
For nuclei heavier than the deuteron no predictions are
available yet. 

In summary, beam--spin asymmetries in the hard electroproduction
of real photons off nuclei have been measured for the first time. 
Sizeable asymmetries of $-0.15 \pm 0.03$~(stat)~$\pm \, 0.03$~(sys) and
$-0.22 \pm 0.03$~(stat)~$\pm \, 0.03$~(sys) have been found in the 
exclusive region for deuterium and neon, respectively.
The corresponding asymmetry on the proton amounts to 
$-0.18 \pm 0.03$~(stat)~$\pm \, 0.03$~(sys). 
This value is in agreement with the already published data from an earlier 
running period.







\end{document}